# SAR-GTR: Attributed Scattering Information Guided SAR Graph Transformer Recognition Algorithm

Xuying Xiong, Xinyu Zhang, Weidong Jiang, *Member, IEEE*, Li Liu *Senior Member*, *IEEE*, Yongxiang Liu, *Member*, *IEEE*, and Tianpeng Liu

*Abstract*—Utilizing electromagnetic scattering information for SAR data interpretation is currently a prominent research focus in the SAR interpretation domain. Graph Neural Networks (GNNs) can effectively integrate domain-specific physical knowledge and human prior knowledge, thereby alleviating challenges such as limited sample availability and poor generalization in SAR interpretation. In this study, we thoroughly investigate the electromagnetic inverse scattering information of single-channel SAR and re-examine the limitations of applying GNNs to SAR interpretation. We propose the SAR Graph Transformer Recognition Algorithm (SAR-GTR). SAR-GTR carefully considers the attributes and characteristics of different electromagnetic scattering parameters by distinguishing the mapping methods for discrete and continuous parameters, thereby avoiding information confusion and loss. Furthermore, the GTR combines GNNs with the Transformer mechanism and introduces an edge information enhancement channel to facilitate interactive learning of node and edge features, enabling the capture of robust and global structural characteristics of targets. Additionally, the GTR constructs a hierarchical topology-aware system through global node encoding and edge position encoding, fully exploiting the hierarchical structural information of targets. Finally, the effectiveness of the algorithm is validated using the ATRNet-STAR large-scale vehicle dataset.

*Index Terms*—SAR ATR, Electromagnetic Scattering Mechanism, Heterogeneous Graph, Attention Mechanism, Feature Fusion, Light Weight.

## I. INTRODUCTION

IN recent years, extensive research has been conducted on physics-informed fusion methods in the field of SAR ATR (Synthetic Aperture Radar Automatic Target Recognition) [1], [2], [3]. The utilization of attributed scattering centers (ASC) information is an indispensable exploration in the interpretation of single-channel SAR data[4], [5], [6]. Multiple studies have demonstrated the effectiveness of incorporating ASC information in enhancing the results of target recognition algorithms [7], [8]. Designing effective methods to utilize

ASC information is crucial for improving the performance of these methods and the generalization capability of the models. Modeling multiple scattering centers in the scene as graph data is an effective approach to combine electromagnetic scattering physical information with the topological geometric representation of the target[9].

Recent studies utilizing graph neural networks[10] (GNNs) to learn invariant properties and geometric features from data, thereby enhancing algorithm performance, have garnered significant attention from researchers in the field[11]. Reference [12] utilizes GNNs to learn structural features of aircraft, addressing the issue of discrete scattering points of aircraft targets in SAR images. Some researchers have combined GNNs with convolutional neural networks (CNNs), where GNNs effectively capture rotation-invariant local structural features, compensating for the limitations of CNNs [13], [14]. Reference [15] designed two methods for extracting graph data nodes to enrich the nodes of the graph and constructed a graph attention network to improve the recognition performance of the algorithm. Reference [16] constructed ASC parameters as heterogeneous graphs, incorporating type information of scattering centers into the graph data. Through multi-head and multi-level feature extraction and learning, this approach achieves better recognition rates under conditions of missing angles and limited samples. The ASC parameters consist of both discrete and continuous values[17]. However, the aforementioned works did not distinguish between parameter types, treating all parameters uniformly. Discrete and continuous values exist in different data spaces and exhibit different distribution characteristics, with distinct statistical properties and semantic meanings[18]. Failing to differentiate between these two types of information in direct mapping and learning may lead to inaccurate feature representations, making it difficult to capture the essential information within the data. Treating discrete information as continuous may introduce unnecessary numerical relationships, while considering continuous values as discrete could result in the loss of important details regarding their numerical variations [16]. To fully extract the semantic features of physical information, this work adopts different mapping methods for different types of parameters, thereby avoiding information loss or confusion.

In addition, the aforementioned GNN constructions are primarily based on Graph Convolutional Networks (GCN) or Message Passing Mechanisms (MPM), but both approaches have certain limitations in the field of Synthetic Aperture

This study was partially supported by the Hunan Provincial Graduate Student Research Innovation Program (CX20240129), the National Key Research and Development Program of China (2021YFB3100800), and the Innovative Research Group Project of the National Natural Science Foundation of China (61921001)" (Corresponding author: Xuying Xiong; Xinyu Zhang).

Xuying Xiong is with the National University of Defense Technology (NUDT), Changsha China (e-mail: xiongxuying@nudt.edu.cn ). Xinyu Zhang is with the NUDT, Changsha China (e-mail: zhangxinyu90111@163.com).
Weidong Jiang is with the NUDT, Changsha China. Li Liu is with the NUDT, Changsha China. Yongxiang Liu is with the NUDT), Changsha China. Tianpeng Liu is with the NUDT, Changsha China.

Radar Automatic Target Recognition (SAR ATR). GCN performs convolution operations using the Laplacian operator of the graph, updating node features based on the adjacency matrix. Therefore, GCN requires that the number of nodes in all input graphs must be the same [19]. However, different categories of targets inherently possess varying numbers of scattering centers. Due to the sensitivity of SAR imaging to observation angles, even targets of the same class can exhibit unequal numbers of scattering center representations from different viewpoints, which directly conflicts with the rigid requirements of GCN. Additionally, GCN relies on two assumptions: the local smoothness assumption (that features of neighboring nodes are similar) and fixed weight sharing (where all edges share the same aggregation weight) [20]. This leads to GCN being unable to handle situations where the features of neighboring nodes exhibit significant differences, as well as difficulty in modeling the dynamic interaction relationships among scattering points. Additionally, GCN is prone to overfitting. The GNN based on the message passing mechanism (MPM) mainly updates node representations by aggregating information from neighboring nodes. This mechanism enables it to adapt to graph data with varying structures [21]. However, this characteristic of local dependence makes it difficult for the model to effectively capture long-range dependencies and global structural features. On one hand, MPM performs poorly in understanding the overall geometric properties of the target, failing to comprehensively reflect the global information of the graph. On the other hand, for nodes that are far apart in the graph, information must be transmitted multiple times to reach its destination, which not only increases computational complexity but may also lead to information loss or degradation during the transmission process.

Transformer mechanism initially achieved great success in the field of natural language processing, and its core self-attention mechanism can simultaneously capture the relationships between any two elements in a sequence without relying on local context [22]. Benefit from its global attention mechanism and dynamic weight allocation capability, Transformer can effectively model long-range dependencies between elements, breaking the limitations of local receptive fields. It can capture global information within a single layer, avoiding delays and degradation in information transmission. Combining the Transformer with GNN extends these advantages further in graph data. Graph Transformer (GT) captures global graph structural information by directly modeling the relationships between any two nodes in the graph through a self-attention mechanism, addressing the limitations of GCN and MPM. GT can also adapt more flexibly to graph with different structures, as the self-attention mechanism allows the model to dynamically select neighbors based on the similarity between nodes. Furthermore, the multi-head attention mechanism of GT can simultaneously capture multiple complex nonlinear relationships, significantly improving the efficiency of feature extraction. Without the need to design complex message-passing mechanisms, GT effectively captures the global features of graph data and demonstrates stronger adaptability and expressive power when handling complex graph structures [23]. Based on considerations of the inductive bias of Transformer, this paper propose for the first time a SAR recognition method that integrates GNN and Transformer, referred to as SAR-GTR. SAR-GTR allows nodes to attend to all other nodes in the graph, enabling the dynamic learning of long-range dependencies between nodes without being constrained by local structures. By jointly learning local geometric structures and global semantic associations, the method significantly enhances the efficiency of ASC information utilization, providing a more robust feature learning paradigm for SAR ATR.

From the perspective of the physical characteristics of SAR data, the edges of a graph represent the correlations between nodes, while the topological structure reflects the relationships between key components of the target. This information is equally important as node attributes, and there exist potential connections among various types of information[24]. Graph data contains not only node information but also the connectivity relationships between nodes. Edge features can describe attributes such as the strength and type of relationships between nodes, making them an essential component of the graph. Relying solely on node features may overlook the diversity of relationships among nodes, while incorporating edge features can provide a more comprehensive representation of the graph structure, enabling the model to capture richer semantic information[25]. By combining edge features, the model can capture higher-order interaction information between nodes, leading to a better understanding of the global structure of the graph. In previous work, we explored methods for constructing edge features but did not incorporate edge information into the feature learning process [16]. In this paper, we further introduce an edge information enhancement channel, effectively controlling computational complexity while enabling interactive learning between node and edge features. In summary, the main innovations and contributions are as follows:

1. We combines GNN with Transformer in the field of SAR ATR for the first time, named SAR-GTR. SAR-GTR not only makes up for the limitations of the traditional GCN and MPM, but also effectively captures the long-distance dependency and global structural characteristics of the target. SAR-GTR avoids information confusion by distinguishing the discrete and continuous information in ASC parameters. A more robust feature learning paradigm is provided for SAR ATR.

2. SAR-GTR creates the edge information enhancement channels to facilitate the learning of interactions between nodes and edge features. This allows the model to better represent graph structure and capture higher-order interactions between nodes, thereby improving the understanding of global target features. The computational complexity is also controlled to ensure the practicality of the method.

3. We constructed a Hierarchical Topology-Aware System (HTAS) using global node coding and edge location coding to fully leverage structural information at different levels of the target. Global node coding obtains node position representations by decomposing the

Laplace matrix, while edge position coding captures edge structural information through random walks. HTAS improves understanding of overall target features and enhances adaptability to complex graph structures.

Section II provides a detailed description of the specific technical details of SAR-GTR. Section III introduces a novel SAR vehicle dataset and validates the effectiveness of the proposed method. Section IV discusses the significance and limitations of the work. Section IV. summarizes the work.

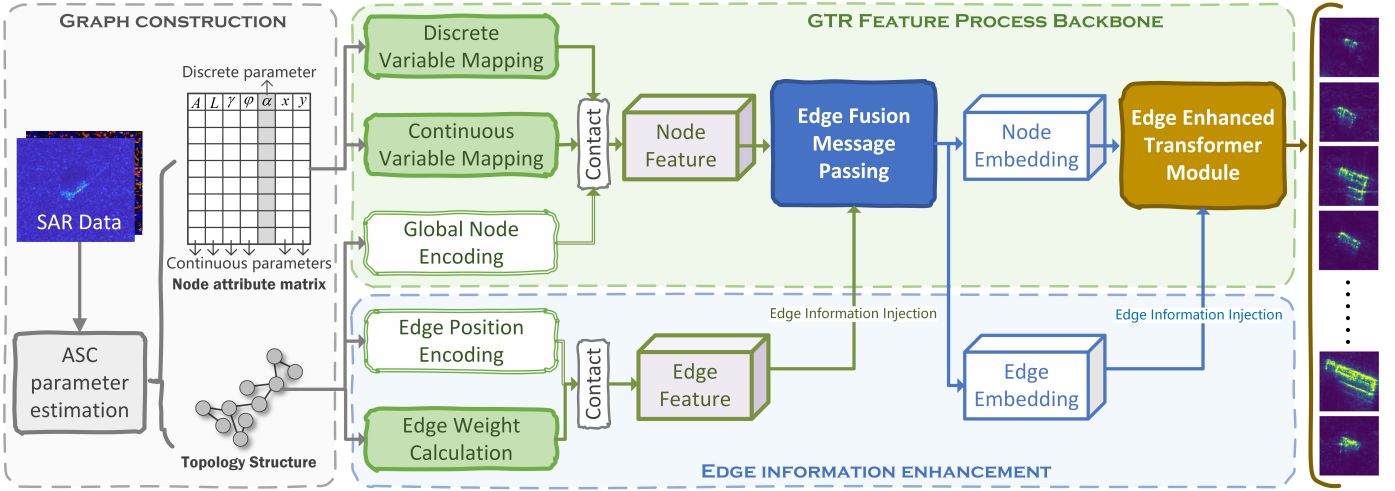

Fig. 1 Overall structure of SAR-GTR and data processing flow diagram.

## II. METHODOLOGY

Radar echo can be regarded as the superposition of multiple scattering centers in the optical region[6]. Assuming that there are a total of $K$ scatterers in the scene, based on the ASC model constructed from single-channel SAR electromagnetic backscatter observations, the $k$th scattering center can be represented by a 7-dimensional parameter vector: $\mathbf{x}_k = \left[ A_k, \alpha_k, L_k, \varphi_k, \gamma_k, x_k, y_k \right]^T \in \mathbb{R}^7$. These parameters reveal the physical structure and characteristics of the scattering center. Due to the imaging mechanism of SAR, the target features exhibit significant variations with changes in squint angles. The graph structure possesses permutation invariance, which effectively addresses angular sensitivity. Therefore, representing the ASC parameters using a graph can fully leverage the domain-specific physical knowledge. The topological structure represented by graphs can better overcome the angular sensitivity inherent in SAR imaging.

Specifically, a graph can be denoted as: $\mathcal{G} = (\mathcal{V}, \mathcal{E}, \mathbf{X})$ , $\mathcal{V} = \{\mathbf{v}_1, \mathbf{v}_2, \ldots, \mathbf{v}_K\}$ is the set of nodes, $\mathcal{E} = \{e_{11}, e_{ij}, \ldots, e_{KK}\}$ is the set of edges. Edge weight is calculated by Gaussian kernel function: $\omega_{ij} = \exp\left(-\|\mathbf{p}_i - \mathbf{p}_j\|_2^2 / 2\sigma_d^2\right) \in \mathbb{R}^+$ ,where $\mathbf{p}_i = (x_i, y_i)$ are the node coordinates, $\sigma_d$ controls the rate at which edge weights decay with real distance between two scatter centers. $\mathbf{H}_0 = \left[\mathbf{h}_1^0, \mathbf{h}_2^0, \ldots, \mathbf{h}_K^0\right]^T \in \mathbb{R}^{K \times 7}$ is the initial attribute matrix, $\mathbf{h}_k^0$ is composed of 7 ASC parameters of the $k$th node. Above graph modeling method maintains the spatial distribution characteristics of scattering centers through the geometric topology relationship, and provides a robust feature representation for SAR ATR.

### A. GTR Overall Model Structure

Based on the graph obtained by the above construction method, we designed GTR to learn the topological and semantic features of electromagnetic scattering information, as illustrated in the Fig. 1. The backbone structure of the GTR is based on the Transformer architecture, which is capable of processing graphs with an arbitrary number of nodes, allowing it to better adapt to different types of target data.

### 1. Discrete Variable Mapping (DVM)

In ASC paraments, only $\alpha$ is discrete, other parameters are continuous. The space of discrete value is different from that of continuous value. Direct feature learning of all parameters can not distinguish the types of parameters, which may lead to confusion of feature space. When a model directly deals with discrete values, it may mistakenly assume that they have a sequential or linear relationship [26]. For this reason, GTR carries out special embedding mapping through discrete variable mapping (DVM) module in the feature embedding process. The mapping of discrete values to low dimensional continuous vector space endows it with dense and learnable semantic representation. On the other hand, the embedded discrete feature and continuous value feature are in the same continuous space, which is convenient for effective feature learning with continuous value feature. The specific steps of DVM are as follows:

TABLE I
DISCRETE PARAMETER MAPPING STEPS

| Steps | Discrete variable mapping process |
|---|---|
| 1 | Map $\alpha$ to integer index: $s_\alpha = \{0,1,2,3,4,5\}$ |
| 2 | Denote a trainable embedding matrix: $\mathbf{W}_{emb}$ |
| 3 | Obtain the embedding vector corresponding to each index: $\mathbf{z}_\alpha = \mathbf{W}_{emb}[s_\alpha]$ |
| 4 | Application of Nonlinear Activation: $\mathbf{h}_\alpha = \mathrm{ReLU}(\mathbf{z}_\alpha)$ |
| 5 | Dimension adjustment. |

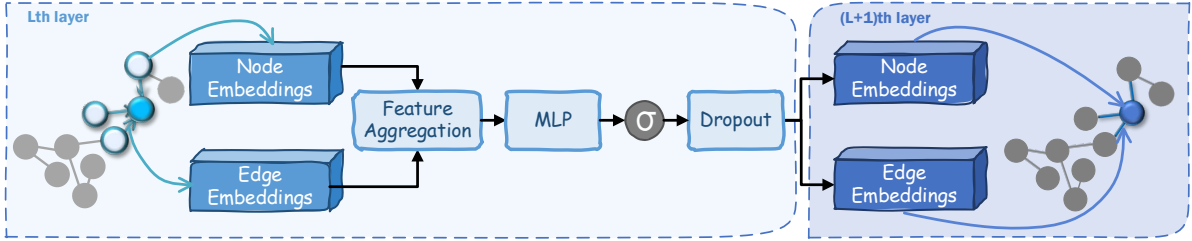

Fig. 2 Edge information enhanced messaging passing mechanism.

The edge channel provides a dedicated processing pathway for edge information. Node embeddings $\mathbf{H} \in \mathbb{R}^{|V| \times d_n}$ is obtained from the initial attribute matrix $\mathbf{H}_0$. Edge embeddings $\mathbf{E} \in \mathbb{R}^{|\varepsilon| \times d_e}$ are derived from the weighted adjacency matrix, which serves as the edge representation. Subsequently, both $\mathbf{H}$ and $\mathbf{E}$ are simultaneously fed into the feature aggregation module in Fig. 2 for feature updating. The update process consists of two steps. In the first step, we calculate the weighted coefficients, which guide the feature generation from node $v_i$ to node $v_j$. This coefficient is essential for determining the influence of node $v_i$ on the feature representation of node $v_j$.

$$\alpha_{ij} = \mathrm{Softmax}_{i \in \mathcal{N}(j)} \left\{ \mathbf{w}_{ij}^T \mathrm{LeakyReLU} \left( \mathbf{W}_{ij} \left( \mathbf{h}_i \| \mathbf{h}_j \| \mathbf{e}_{ij} \right) \right) \right\} \quad (1)$$

where $\mathbf{W}_{ij} \in \mathbb{R}^{d \times (2d_n + d_e)}$ and $\mathbf{w}_{ij} \in \mathbb{R}^d$ are both trainable parameters, $\mathcal{N}(j)$ denotes the set of neighbor nodes of $v_j$. $\|$ means vector contact. Based on $\alpha_{ij}$ with edge information, the new feature $\mathbf{h}_j'$ of node $v_j$ can be get by aggregating embedding of all neighbor nodes. While the new edge feature $\mathbf{e}_{ij}'$ is obtained by residual update:

$$\begin{cases} \mathbf{h}_j' = \sum_{i \in \mathcal{N}(j)} \alpha_{ij} \cdot \mathbf{W}_{agg} \mathbf{h}_i \\ \mathbf{e}_{ij}' = \mathbf{e}_{ij} + \mathbf{W}_e \left( \alpha_{ij} \cdot \mathbf{e}_{ij} \right) \end{cases} \quad (2)$$

## 2. Edge Information Enhancement

DVM alleviates the issues of sparsity and semantic loss associated with discrete values, preventing erroneous handling of these values and thereby enhancing feature learning capability of GTR. To fully exploit the potential semantic information from all levels and types of graph and strengthen the ability of GTR to learn target structural features. Unlike traditional GNNs that focus solely on node features, GTR integrates edge information enhancement channels in both the Transformer and MPM, facilitating comprehensive interaction between structural and node information. The schematic structure of the MPM module after incorporating edge information channels is illustrated in Fig. 2.

where $\mathbf{W}_{agg} \in \mathbb{R}^{d_n \times d_n}$ and $\mathbf{W}_e \in \mathbb{R}^{d_e \times d_e}$ are the trainable parameter matrix of node aggregation and edge update, respectively. The residual connection mechanism effectively retains the original edge features to ensure the stability of model training.

Transformer structure with the inclusion of an edge information enhancement channel is shown in Fig. 3. In classical Transformer architecture, the processing typically only involves node features, specifically generating the $\mathbf{Q}$ $\mathbf{K}$ $\mathbf{V}$ matrix solely from node embeddings. However, the flexible architecture of the transformer allows for the incorporation of edge information. After modifying the structure, we utilize edge embeddings as an additional source for generating the $\mathbf{Q}$ matrix.

Reorganize all node features into central node feature matrix $\mathbf{H}_s$ and neighbor node feature matrix $\mathbf{H}_i$ according to edge index $e_{ij} = (v_i, v_j)$. $v_j$ represents the central node, $v_i$ is the neighbor node. After a simple linear layer mapping transformation, $\mathbf{V} = \mathrm{MLP}(\mathbf{W}_{vn} \cdot \mathbf{H}_i)$, $\mathbf{K} = \mathrm{MLP}(\mathbf{W}_k \cdot \mathbf{H}_i)$, $\mathbf{Q}_n = \mathrm{MLP}(\mathbf{W}_{qn} \cdot \mathbf{H}_s)$, $\mathbf{Q}_e = \mathrm{MLP}(\mathbf{W}_{qe} \cdot \mathbf{E})$, $\mathbf{V}_e = \mathrm{MLP}(\mathbf{W}_{qe} \cdot \mathbf{E})$ can be obtained. $\mathbf{Q}_n$ is generated by the central node, indicating what information the central node needs to obtain from its neighbors. $\mathbf{V}$ is generated by the embedding of neighbor nodes and represents the content that neighbor nodes can provide. $\mathbf{Q}_e$ is the query matrix generated by edge embedding, which participates in the calculation of attention score and uses edge features to guide the flow of information. $\mathbf{V}_e$ directly acts on the feature generation stage, preserving the edge semantic information and enhancing the information content of the output feature.

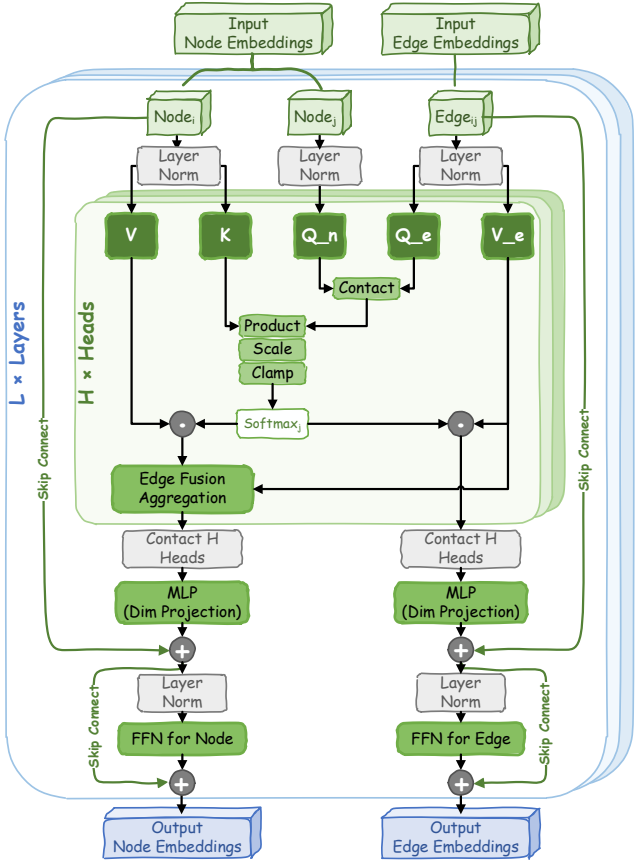

Fig. 3 Structure of edge information enhanced graph transformer module.

First, we use $\mathbf{Q}_n$ and $\mathbf{Q}_e$ to calculate the attention weight:

$$s_{ij} = \text{Softmax}_{i \in \mathcal{N}(j)} \left( \frac{(\mathbf{q}_n^{(i)} \| \mathbf{q}_e^{(ij)}) \mathbf{W}_s \cdot \mathbf{k}^{(j)}}{\sqrt{d_h}} \right) \quad (3)$$

The filtering weight of edge features and requirements of nodes pairs are encoded by $s_{ij}$ at the same time. Then use $s_{ij}$ to know the generation of node and edge features:

$$\begin{cases} \mathbf{e}_{ij} = \alpha_{ij} \mathbf{v}_e^{(ij)} \\ \mathbf{z}_n^{(j)} = \sum_{i \in \mathcal{N}(j)} s_{ij} \cdot \left( \mathbf{v}_n^{(i)} \| \mathbf{v}_e^{(ij)} \right) \mathbf{W}_z \end{cases} \quad (4)$$

In each layer of the transformer, multiple attention heads are employed, allowing different heads to focus on various pieces of information from different feature subspaces, thereby enriching the semantic meaning of the output features. Each attention head achieves the decoupling and complementary feature space through independent parameterization. After concatenating the node and edge features obtained from (4) for each attention head, the input for the next layer is generated through dimensional mapping and normalization. The introduction and processing of edge information enhance the expressiveness of model, enabling a more in-depth exploration of the hierarchical features within graph-structured data.

### B. Hierarchical Topology-Aware System (HTAS)

#### 1. Global Node Encoding(GNE)

To enhance the ability to express graph topological structures, GTR innovatively introduces two key components: the Global Node Encoder (GNE) and the Edge Position Encoding (EPE). These components work in synergy to comprehensively capture the topological features of the graph. GNE adopts the principles of spectral graph theory, leveraging the global topological information and local connectivity relationships contained in the Laplacian matrix to establish a structure-aware feature representation for each node. Specifically, GNE performs spectral decomposition on the graph Laplacian matrix to extract feature vectors that serve as the positional encoding for the nodes. The normalized Laplacian matrix of the graph is defined as follows:

$$\mathbf{L} = \mathbf{I} - \mathbf{D}^{-1/2} \mathbf{A} \mathbf{D}^{-1/2} \quad (5)$$

where $\mathbf{A} \in \mathbb{R}^{K \times K}$ is the adjacency matrix without edge weight, $\mathbf{D} \in \mathbb{R}^{K \times K}$ is the degree matrix, $D_{ii} = \sum_{j=1}^{K} A_{ij}$. By decomposing $\mathbf{L} = \mathbf{Q} \mathbf{\Lambda} \mathbf{Q}^T$, the eigenvector matrix $\mathbf{Q} = [\mathbf{q}_1, ..., \mathbf{q}_K] \in \mathbb{R}^{K \times K}$ and eigenvalue matrix $\mathbf{\Lambda} = \text{diag}(\lambda_1, ..., \lambda_K)$ can be obtained. $\mathbf{q}_k \in \mathbb{R}^{1 \times d_{qk}}$ corresponds to the spectral embedding representation of the $k$th node on the graph, and the eigenvalue $\lambda_k$ measures the importance of the corresponding eigenvector in the graph topology. In order to achieve a balance between computational efficiency and expression ability, only the eigenvectors corresponding to the first $n$ minimum eigenvalues (sorted by $\lambda_1 \leq \lambda_2 \leq ... \leq \lambda_n$) are selected in practical applications. The position encoding of node $v_k$ can be write as:

$$\text{GNE}(v_k) = [q_{k1}, q_{k2}, \cdots, q_n] \in \mathbb{R}^{1 \times n} \quad (6)$$

where $q_{ki}$ is the $l$th elements of $\mathbf{q}_k$. GNE is based on the important properties of spectral graph theory [27]: smaller eigenvalues and their corresponding eigenvectors (low-frequency components) more accurately reflect the global connectivity and macrostructural characteristics of the graph, while larger eigenvalues correspond to high-frequency components that may contain more local details or noise. The macrostructural features are of greater interest in the SAR ATR task. GNE not only effectively characterizes the topological positional relationships of nodes within the graph but also maintains permutation invariance, which is independent of the arrangement order of the nodes. This provides a structure-aware node representation that enhances the ability of model to extract graph features.

### 2. Edge Position Encoding (EPE)

EPE aids the model in better capturing the structural information of edges and the relative positional relationships of nodes, thereby enhancing the ability to discern connection patterns. To prevent information loss due to insufficient prior knowledge, we construct the ASC parameters as a fully connected graph. However, to avoid excessive computational costs associated with encoding in a fully connected dense graph, EPE employs a random walk-based encoding approach[28]. In a weighted graph, edges with high weights typically correspond to crucial connections; thus, their high encoding values can assist the model in distinguishing between important and non-important edges. EPE integrates edge weight information into the encoding process by dynamically adjusting the transition probabilities of the random walk. Specifically, the relationship between the probability of transition from $v_i$ to $v_j$ and its edge weight $\omega_{ij}$ is:

$$p(v_j \mid v_i) = \frac{\omega_{ij}}{\sum_{k \in \mathcal{N}(i)} \omega_{ik}} \tag{7}$$

Introducing weights into the transition probabilities allows the random walk paths to preferentially propagate along edges with higher weights. For an undirected weighted graph, which means $\omega_{ij} = \omega_{ji}$, the stationary distribution $\pi_i \in \mathbb{R}^K$ of the random walk Markov chain is defined as follows:

$$\pi_i = \frac{\sum_{j \in \mathcal{N}(i)} \omega_{ij}}{\sum_{k=1}^{K} \sum_{l \in \mathcal{N}(k)} \omega_{kl}} = \frac{d_i}{2W} \tag{8}$$

where $d_i = \sum_{j \in \mathcal{N}(i)} \omega_{ij}$ is the weighted degree of $v_i$. The sum of edge weights of the whole graph is $W = \frac{1}{2} \sum_{k=1}^{K} \sum_{l \in \mathcal{N}(k)} \omega_{kl}$. $\pi_i$ represents the probability of staying at $v_i$ after a long-term random walk, which is determined by the connection weights between $v_i$ and its neighboring nodes. In the stationary state, the expected visit count of $e_{ij}$ is determined by the stationary probabilities of its endpoints and the transition probabilities. Let the entire graph undergo $N_w$ random walks, and let the average length of a single random walk be $l_w$. The total visit count of $e_{ij}$ is given by:

$$\mathbb{E}[\text{count}(e_{ij})] = N_w l_w \cdot \left( \pi_i p(v_j \mid v_i) + \pi_j p(v_i \mid v_j) \right) = N_w l_w \cdot \frac{\omega_{ij}}{W} \tag{9}$$

Equation (9) indicates that the visit frequency of edges in the EPE is strictly proportional to the edge weight $\omega_{ij}$. The encoding results integrate the structural information of the graph through the paths covered by the random walks and fuse the semantic information through the edge weights. Despite the introduction of weight calculations, the time complexity remains $\mathcal{O}(N_w K l_w)$, indicating suitability for dense weighted graphs. Based on the aforementioned definitions, the final mathematical form of the EPE encoding is:

$$\text{EPE}(e_{ij}) = \frac{\text{count}(e_{ij})}{\sum_{e_{kl} \in \varepsilon} \text{count}(e_{kl})} \in \mathbb{R} \tag{10}$$

where $\text{count}(e_{ij}) = N_w l_w \omega_{ij} / 2W$ represents the expected frequency of $e_{ij}$ being visited across all random walk process. EPE modulates the transition probabilities of the random walks through the edge weights, accurately modeling the strength of edge semantics while preserving global structural information. If the edge weights are updated during training, EPE can resample the walk paths only for the edges with weight changes and their neighborhoods, locally updating the statistics without the need for a complete recompilation of the graph.

The encoding mechanisms for nodes and edges complement each other, with GNE ensuring a comprehensive understanding of the graph structure at a macro level, while EPE enhances the representation of local connectivity details at a micro level, together forming a hierarchical topology-aware system. This design not only enables the model to deal with arbitrary structure graph data adaptively, but also significantly improves the expression ability and generalization performance of the model while maintaining the invariance of the graph.

## III. EXPERIMENTS

### A. Description of ATRNet-STAR dataset

The renowned MSTAR dataset has become insufficient for validating the performance of current target recognition algorithms due to its scale and idealized collection conditions, with many algorithms having already achieved peak performance on MSTAR. In 2025, the CSA Laboratory of National University of Defense Technology constructed a large-scale vehicle SAR dataset named ATRNet-STAR [29]. This dataset was collected using a UAV platform in strip mode and includes HH, HV, VH, and VV polarization data under X-band (9.6 GHz) and Ku-band (14.6 GHz). Both bands have a bandwidth of 1200 MHz, with a measured range resolution of 0.148m and an azimuth resolution of 0.123m. The collection pitch depression angles for all scenes are 15°, 30°, 45°, and 60°. The azimuth squint angle ranges from 0° to 360°, with slight differences in the azimuth angle collection intervals for different scenes, as shown in Table II.

TABLE II
ACQUISITION PARAMETERS FOR DIFFERENT SCENES OF
ATRNET-STAR

| Scene | Class | Band | Azimuth Interval | Amplitude Data | Complex Data |
|-------|-------|------|------------------|----------------|--------------|
| Urban | 40 | X | 5° | 73956 | 74333 |
| Factory | 40 | X, Ku | 30° | 35654 | 33653 |
| Sandstone | 40 | X, Ku | 30° | 16512 | 16512 |
| Woodland | 14 | X, Ku | 30° | 5293 | 5293 |
| Baresoil | 14 | X, Ku | 30° | 5376 | 5376 |

The ATRNet-STAR dataset encompasses 40 different vehicle targets across 5 scenes, with detailed information on target size distribution illustrated in Fig. 4. The vehicle targets in ATRNet-STAR can be categorized into 4 major classes, 21 sub-classes, and 40 specific models, covering most categories of civilian vehicles. Additionally, the dataset provides comprehensive annotation information, including 8-bit unsigned integer amplitude images and 32-bit floating-point complex data, making it the largest publicly available vehicle SAR dataset to date, being 10 times larger than the MSTAR dataset [30]. ATRNet-STAR imposes higher demands on algorithm performance, enabling a more comprehensive evaluation of algorithm effectiveness.

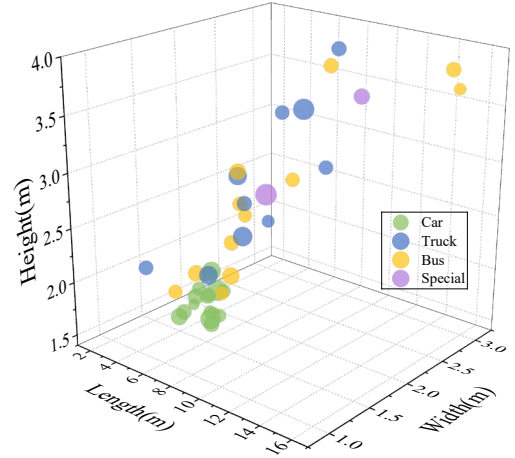

Fig. 4 ATRNet-ATR data distribution scatter plot. The colors represent the major categories of targets, while the radius of the circles indicates the volume of data. The coordinates of the circles correspond to the sizes of the vehicle targets.

This paper follows the seven experimental scenarios provided by the CSA Laboratory for algorithm validation. The specific settings for the experimental scenarios are shown in the Table III. SOC-50 is an experimental scenario comprised of 10 vehicle target categories augmented by the MSTAR. To mitigate the long-tail effect, SOC-50 restricts the number of samples per category to approximately 720 images.

TABLE III
SEVEN EXPERIMENTAL SCENARIOS OFFICIALLY ANNOUNCED BY ATRNET STAR

| No. | Setting | Set | Scene | Depression(°) | Azimuth(°) | Band | Polarization | Quantity |
|-----|---------|-----|-------|---------------|------------|------|--------------|----------|
| 1 | SOC-40 | Train | All | 15,30,45,60 | 0~360 | X, Ku | quad | 67780 |
| | | Test | All | 15,30,45,60 | 0~360 | X, Ku | quad | 29169 |
| 2 | SOC-50 | Train | All | 15,17,30,45,60 | 0~360 | X, Ku | quad | 18071 |
| | | Test | All | 15,30,45,60 | 0~360 | X, Ku | quad | 17613 |
| 3 | EOC-Scene | Train | **Baresoil, Sandstone** | 15,30,45,60 | 0~360 | X, Ku | quad | 19584 |
| | | Test | **City,Factory,Woodland** | 15,30,45,60 | 0~360 | X, Ku | quad | 77365 |
| 4 | EOC-Depression | Train | All | **15** | 0~360 | X, Ku | quad | 22206 |
| | | Test | All | **30,45,60** | 0~360 | X, Ku | quad | 74743 |
| 5 | EOC-Azimuth | Train | All | 15,30,45,60 | **0~60** | X, Ku | quad | 18592 |
| | | Test | All | 15,30,45,60 | **60~360** | X, Ku | quad | 78357 |
| 6 | EOC-Band | Train | All | 15,30,45,60 | 0~360 | **X** | quad | 27653 |
| | | Test | All | 15,30,45,60 | 0~360 | **Ku** | quad | 27732 |
| 7 | EOC-Polarization | Train | All | 15,30,45,60 | 0~360 | X, Ku | **HH** | 24246 |
| | | Test | All | 15,30,45,60 | 0~360 | X, Ku | **other** | 72703 |

## B. Comparative Experiment

Our model is applied on PyTorch. The ASC estimation algorithm is built based on MATLAB 2023b. The CPU and GPU used in the experiment are the 13th generation Intel(R) Core(TM) i9-14900K and NVIDIA GeForce RTX 4090, with a capacity of 24 GB.

Three indicators need to be specified in advance when extracting ASC parameters: minimum peak level, maximum number of scattering centers and maximum fitting percentage. Among them, the maximum fitting percentage has the strongest constraint ability and is also the most important indicator. We first find the best ASC parameter extraction setting through a set of pre-experiments. Specifically, we investigated the probability of correct classification (PCC) under different maximum fitting percentages. The experimental results are shown in the Fig. 5. Thus, the maximum fitting percentage of ASC parameters extraction in ATRNet-STAR is set to 92%. According to empirical parameters in [16], minimum peak level is set to −20 dB and maximum number of scattering centers is set to 40.

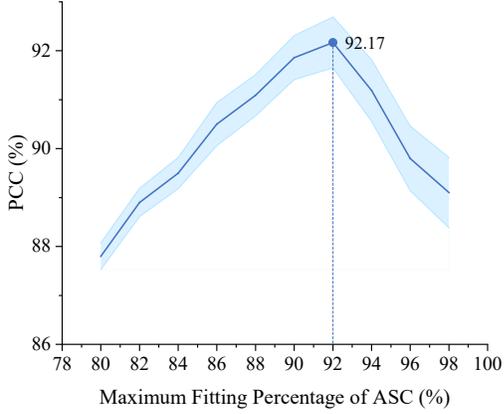

Fig. 5 PCCs of SAR-GTR at different maximum fitting percentages. The error bande is obtained by plotting the results of 10 experiments.

The comparison algorithms are chosen as follows. GAT[31] serves as a foundational graph attention mechanism. GCN[19] is a classic graph convolutional algorithm. ST-Net[12] constructs graphs on amplitude images and combines topological features with image features extracted by CNNs in a dual-stream approach. MSGCN[14] is a dual-stream algorithm that leverages both GCN and CNN, with graph data originating from SAR complex data. VSFA[15] improves the method of graph construction, accomplishing classification tasks using a single GNN model. LDSF[16] is a dual-stream algorithm that combines heterogeneous graph attention network and SE-ResNet18[32]. MS-CVNet[33] is an algorithm that directly utilizes complex data, which serves as a reference benchmark method for the way complex information is utilized. VGG16[34] serves as a reference method for classical classification models. The best result is marked in red, and the second best result is marked in blue. The results are the averages from 10 experiment.

TABLE IV

PCC OF GNN MODEL. FOR THE TWO-WAY ALGORITHM IN THE TABLE, ONLY GNN MODULE IS USED FOR EXPERIMENT

| GNN Model Setting | MS-CVNet | VGG16 | GAT | GCN | ST-Net | MSGCN | VSFA | LDSF | GTR |
|---|---|---|---|---|---|---|---|---|---|
| SOC-40 | 80.91 | 88.93 | 91.80 | 89.94 | 89.72 | 90.49 | 91.43 | 91.85 | 92.17 |
| SOC-50 | 55.23 | 72.82 | 71.07 | 71.90 | 71.74 | 71.54 | 71.41 | 73.87 | 74.61 |
| EOC-Scene | 26.11 | 16.61 | 24.47 | 23.09 | 21.56 | 26.44 | 16.98 | 30.52 | 31.16 |
| -city | 26.06 | 22.74 | 24.04 | 22.33 | 20.22 | 28.48 | 19.14 | 32.00 | 33.38 |
| -factory | 25.59 | 20.32 | 25.40 | 26.13 | 24.98 | 26.12 | 16.56 | 31.63 | 32.69 |
| -woodland | 20.24 | 6.78 | 23.96 | 20.82 | 19.47 | 24.73 | 15.25 | 27.92 | 27.40 |
| EOC-Depression | 21.78 | 33.29 | 33.61 | 33.69 | 37.22 | 35.05 | 35.05 | 36.60 | 37.14 |
| -30° | 39.21 | 52.42 | 51.92 | 52.20 | 56.49 | 52.33 | 52.23 | 55.18 | 56.37 |
| -45° | 19.17 | 33.39 | 33.66 | 33.91 | 38.05 | 33.85 | 37.10 | 38.04 | 38.12 |
| -60° | 8.07 | 14.05 | 15.24 | 14.96 | 17.13 | 14.32 | 15.83 | 16.57 | 16.94 |
| EOC-Azimuth | 19.03 | 15.90 | 22.43 | 22.41 | 20.46 | 23.04 | 25.06 | 25.71 | 26.09 |
| -60° | 26.82 | 20.84 | 32.42 | 30.61 | 29.84 | 29.86 | 34.42 | 34.26 | 34.98 |
| -120° | 13.24 | 8.01 | 17.31 | 18.15 | 14.84 | 17.71 | 19.59 | 20.05 | 20.00 |
| -180° | 16.89 | 17.61 | 20.15 | 20.44 | 17.83 | 21.01 | 20.84 | 22.26 | 22.85 |
| -240° | 10.94 | 7.22 | 16.97 | 16.82 | 15.40 | 17.15 | 19.27 | 19.52 | 19.83 |
| -300° | 28.77 | 25.81 | 25.29 | 26.01 | 24.41 | 29.48 | 31.20 | 32.44 | 32.78 |
| EOC-Band | 63.73 | 78.95 | 81.81 | 82.33 | 81.23 | 83.44 | 83.27 | 83.25 | 84.77 |
| -inverse | 60.92 | 74.37 | 75.06 | 74.70 | 74.98 | 78.41 | 78.08 | 76.33 | 79.31 |
| EOC-Polarization | 49.03 | 72.42 | 70.63 | 70.67 | 68.75 | 70.09 | 70.60 | 70.41 | 71.05 |
| -VV | 52.01 | 72.23 | 71.78 | 71.48 | 69.06 | 70.14 | 71.28 | 72.06 | 72.07 |
| -HV | 45.78 | 72.31 | 70.44 | 70.98 | 69.20 | 70.83 | 70.78 | 70.06 | 71.16 |
| -VH | 46.12 | 72.71 | 69.67 | 69.56 | 68.00 | 69.94 | 69.76 | 69.12 | 69.93 |

The original versions of ST-Net, MSGCN and LDSF are dual-stream algorithms, but the results in Table IV are from the GNN branches only. Table IV aims to examine the classification ability of the GNN model alone. It can be seen that SAR-GTR has a clear superiority among all GNN models. MS-CVNet directly uses complex information but performs the worst, indicating that while the raw data contains the most complete information about the target, the method of utilizing that information significantly impacts the final classification results. In EOC-woodland scenario, LDSF performs best. In this scenario, trees in the background cause more pronounced layering and multipath effects. As Fig. 6(b) shows, more background scattering information is mixed in the scattering cells corresponding to the vehicle target, and the contamination of the ASC parameters is more obvious. Moreover, the scattering performance of trees usually shows as discrete centers. While LDSF enhances the focus on the characteristics of target by distinguishing the types of the scattering centers. In strong background clutter, this approach allows for a more accurate representation of the structural features of targets, leading to better results. In EOC-Depression scenario, GTR has almost the same effect as ST-Net. As can be seen in Fig. 6(c), depression angle variation leads to significant differences in the number of backscatter centers in SAR images. By fixing the number of scattering centers to 9, ST-Net counteracts to some extent the feature confusion caused by fluctuations in the number of scattering centers.

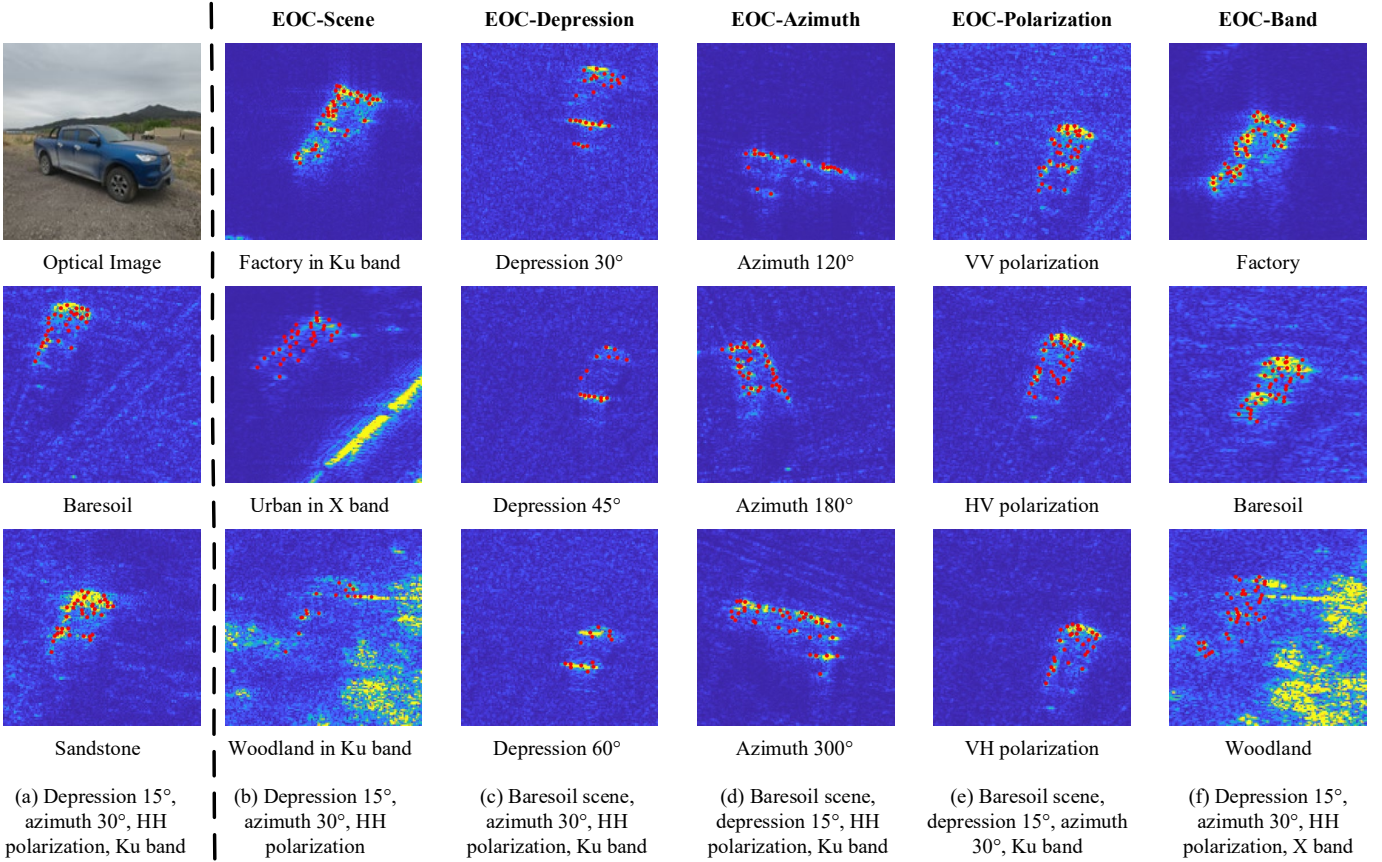

Fig. 6 Magnitude data visualization of targets named Great_Wall_poer in ATRNet-STAR. The red points in the figure indicate the extracted scattering centers. The left side of the dotted line is the training data, and the right side corresponds to the test data in different EOC scenarios.

In EOC-Azimuth scenario, different angles lead to geometries that exhibit different backscattering. Reflected in the ASC parameter, it is the value of $\alpha$ that produces a noticeable change. SAR-GTR distinguishes the attribute of $\alpha$ by DVM, which effectively preserves the information of discrete parameters. In addition, the edge information enhancement and the effective characterization of the structure of the graph data by HTAS enable SAR-GTR to better overcome the classification dilemma caused by SAR angle sensitivity. In EOC-Polarization scenario, VGG16 has the best performance. From Fig. 7, we can see that changes in polarization conditions mainly affect scattering properties, which are reflected in visual effects such as texture and intensity. Therefore, the VGG model, which is more adapted to image features, performs better than the GNN.

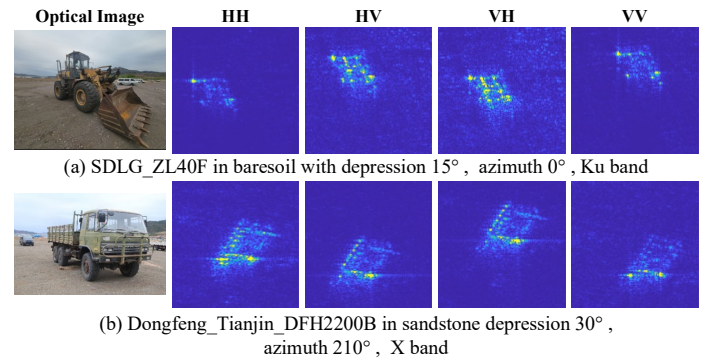

(a) SDLG_ZL40F in baresoil with depression 15°, azimuth 0°, Ku band

(b) Dongfeng_Tianjin_DFH2200B in sandstone depression 30°, azimuth 210°, X band

Fig. 7 Comparison of SAR amplitude images under different polarization conditions. Each slice of data carries a different degree of offset.

Table V shows the PCC in dual-stream mode, which classifies features extracted by fusing GNN and CNN models. In this case, the CNN module is used as a branch for complementary visual features, while GAT, GCN, VSFA and GTR are all experimented using the GVF-Net module in LDSF as the CNN branch. From Table V, it can be seen that the overall performance of the algorithms using the dual-stream architecture is better than that of the single GNN method, which indicates that GNN and CNN focus on different and complementary target features, and the combination of the two can significantly improve the performance of the algorithms. Among all dual-stream methods, GTR still leads the performance, especially in the EOC-Depression and EOC-Polarization scenarios. This proves that the effectiveness of the GNN module remains the basis of the algorithm's performance, despite the CNN can provide support for visual information. In several subclasses of the EOC-Depression scenario, SAR-GTR outperforms ST-Net, and the inclusion of visual features effectively mitigates the negative impact of the variation in the number of scattering points. In the EOC-Polarization scenario, by supplementing visual information such as texture, GTR surpasses VGG16 and achieves the best classification results.

TABLE V
PCC OF DUAL-STREAM ALGORITHMS. GAT, GCN AND SAR-GTR ALL USE SEResNet-18 AS ANOTHER WAY TO SUPPLEMENT VISUAL FEATURES

| Dual-Stream Setting | MS-CVNet | VGG | GAT | GCN | ST-Net | MSGCN | VSFA | LDSF | GTR |
|---|---|---|---|---|---|---|---|---|---|
| SOC-40 | 80.91 | 88.93 | 92.51 | 91.73 | 91.38 | 92.26 | 92.14 | 92.62 | 93.25 |
| SOC-50 | 55.23 | 72.82 | 73.13 | 72.73 | 73.01 | 73.12 | 72.38 | 74.51 | 75.09 |
| EOC-Scene | 26.11 | 16.61 | 25.75 | 24.37 | 23.93 | 27.50 | 17.83 | 31.09 | 31.77 |
| -city | 26.06 | 22.74 | 25.35 | 23.52 | 23.04 | 29.44 | 20.44 | 32.74 | 34.03 |
| -factory | 25.59 | 20.32 | 27.14 | 27.09 | 26.76 | 27.23 | 17.23 | 32.18 | 33.37 |
| -woodland | 20.24 | 6.78 | 24.77 | 22.51 | 21.99 | 25.83 | 15.83 | 28.35 | 27.92 |
| EOC-Depression | 21.78 | 33.29 | 35.87 | 34.46 | 37.68 | 34.74 | 36.09 | 36.98 | 37.75 |
| -30° | 39.21 | 52.42 | 54.15 | 52.74 | 56.58 | 53.02 | 53.36 | 55.26 | 56.49 |
| -45° | 19.17 | 33.39 | 36.11 | 34.70 | 38.89 | 34.98 | 37.77 | 38.22 | 39.18 |
| -60° | 8.07 | 14.05 | 17.36 | 15.95 | 17.57 | 16.23 | 17.15 | 17.47 | 17.59 |
| EOC-Azimuth | 19.03 | 15.90 | 24.06 | 23.89 | 22.93 | 24.25 | 26.15 | 26.37 | 27.14 |
| -60° | 26.82 | 20.84 | 32.97 | 32.56 | 30.84 | 31.71 | 35.59 | 35.09 | 35.77 |
| -120° | 13.24 | 8.01 | 19.28 | 19.07 | 18.15 | 19.13 | 20.31 | 20.63 | 22.26 |
| -180° | 16.89 | 17.61 | 21.31 | 21.90 | 20.18 | 21.96 | 22.54 | 22.71 | 23.06 |
| -240° | 10.94 | 7.22 | 19.33 | 17.92 | 18.20 | 18.47 | 20.35 | 19.65 | 20.95 |
| -300° | 28.77 | 25.81 | 27.39 | 27.98 | 27.26 | 29.98 | 31.94 | 33.75 | 33.68 |
| EOC-Band | 63.73 | 78.95 | 84.04 | 83.63 | 83.91 | 84.37 | 84.25 | 84.15 | 85.22 |
| -inverse | 60.92 | 74.37 | 77.18 | 76.61 | 76.89 | 79.55 | 79.43 | 77.13 | 80.13 |
| EOC-Polarization | 49.03 | 72.42 | 72.31 | 72.23 | 72.51 | 72.31 | 71.88 | 72.88 | 73.32 |
| -VV | 52.01 | 72.23 | 73.85 | 73.44 | 73.72 | 72.62 | 71.59 | 73.96 | 74.43 |
| -HV | 45.78 | 72.31 | 71.16 | 71.75 | 72.03 | 72.03 | 71.91 | 72.47 | 72.89 |
| -VH | 46.12 | 72.71 | 71.91 | 71.50 | 71.78 | 72.27 | 72.15 | 72.22 | 72.64 |

In addition, during the training process, we found that LDSF needs to contain both distributed and discrete scattering centers, but this condition is difficult to satisfy in some scenes, so the application range of this algorithm is somewhat limited. Meanwhile, GAT and GCN are prone to overfitting phenomenon during the training process, and the hyperparameters need to be carefully selected to ensure the stable training of the model.

*C. Ablation Experiment*

To validate the effectiveness of each module in the algorithm, this section designs an ablation study, the results of which are presented in the following table. During the testing phase, the EOC experimental scenario employs all subsets of data for a single test without further subdividing the scenes.

Table VI shows the PCC of SAR-GTR in each experimental scenario with different modules missing, where None represents the full GTR model and Dual represents the model after combining GTR and GVF-Net in LDSF. It can be seen that the dual-stream model achieves the best performance in all scenarios. However, the full GTR model achieved the second best results in all scenarios except EOC-Polarization, the model with missing DVM instead outperforms the full GTR model in EOC-Polarization. Different polarization conditions affect several echo attributes such as scattering intensity, scattering coefficient, phase information, scattering mechanism, and echo signature of the target. While the ASC model is mainly applicable to SAR data under single polarization conditions. Using a single-polarization model to deal with multi-polarized data is likely to introduce errors in parameter estimation. DVM is depend on the properties of the ASC parameters, so the effectiveness of the module needs to be based on the premise that the ASC parameters are extracted accurately. The practice of DVM that subdivide the ASC parameters is very likely to amplify the different polarization features, which leads to the model's confusion in the category judgment. In dual-stream, the performance of the model is obviously improved after the target feature information is supplemented by the image feature extraction module from the perspective of contour and texture.

TABLE VI
PCC AFTER REMOVING DIFFERENT MODULES IN SAR-GTR

| Ablation Setting | DVM | Edge-Enhanced | GPE | EPE | None | Dual |
|---|---|---|---|---|---|---|
| SOC-40 | 92.00 | 91.80 | 92.07 | 91.93 | 92.17 | 93.25 |
| SOC-50 | 73.75 | 73.56 | 73.93 | 73.76 | 74.61 | 75.09 |
| EOC-Scene | 30.48 | 30.29 | 30.57 | 30.47 | 31.16 | 31.77 |
| EOC-Depression | 36.29 | 36.16 | 36.40 | 36.34 | 37.14 | 37.75 |
| EOC-Azimuth | 25.24 | 25.00 | 25.39 | 25.32 | 26.09 | 27.14 |
| EOC-Band | 83.92 | 83.71 | 84.05 | 84.00 | 84.77 | 85.22 |
| EOC-Polarization | 72.49 | 70.55 | 70.76 | 69.88 | 71.07 | 73.32 |

## IV. DISCUSSION

Choosing GNNs to utilize ASC for SAR ATR is a method that effectively integrates physical information with target topology. In this paper, we explore how to leverage a more rational GNN model to fully exploit the topological structure features of target and the semantic information contained within physical knowledge. Distinguishing the parametric characteristics within ASC parameters is largely unaffected by the data acquisition scene conditions and can significantly improve the performance of model.

To address the issue of overfitting commonly observed in foundational GNN models, we investigate their underlying computational principles. We find that GCNs and MPMs struggle to capture the global information of the graph during computation. This limitation hampers effective model convergence and restricts feature learning capabilities. Therefore, employing a Transformer-based approach to enhance the model architecture can effectively mitigate these drawbacks. The proposed HTAS facilitates the injection of structural information from the graph into the model. It is noteworthy that the proposed GNE and EPE schemes impose no structural requirements on the model and can be used to improve the performance of any traditional GNN by providing additional global information.

Finally, similar to all current SAR ATR algorithms utilizing ASC information, SAR-GTR is constrained by the efficiency of ASC parameter extraction. In the single-channel SAR ATR domain, there is an urgent need to investigate intelligent extraction algorithms for ASC parameters to achieve end-to-end ATR. Additionally, there currently exists no dataset to support electromagnetic inverse scattering studies for single-channel SAR data. This necessitates collaborative efforts from industry professionals to contribute to research in this direction.

## V. CONCLUSION

Effectively integrating domain-specific physical knowledge with the target topology is a critical issue that needs to be addressed in current SAR ATR tasks. We re-examine the intrinsic properties of electromagnetic scattering parameters and carefully analyze the limitations of existing GNN algorithms. Considering the characteristics of the SAR ATR task, we propose the SAR-GTR model. The SAR-GTR model is designed to align more closely with the essential features of the target in terms of data structure and topological characteristics, allowing for the learning of more robust classification features. Finally, we conduct comprehensive testing of the algorithm on the latest ATRNet-STAR dataset, validating the effectiveness of the proposed approach..